\documentclass[aps,twocolumn,prl,tightenlines,floatfix,showpacs]{revtex4}

\usepackage[dvips]{graphicx}
\usepackage[english]{babel}
\usepackage{amsmath}
\usepackage{amssymb} 
\usepackage{slashed}
\begin{document}

\title{Microscopic Approach to Shear Viscosities in Superfluid Gases:
From BCS to BEC}

\author{Hao Guo $^{1}$, Dan Wulin $^{1}$,
Chih-Chun Chien$^{2}$ and K. Levin$^{1}$}

\affiliation{$^1$James Franck Institute and Department of Physics,
University of Chicago, Chicago, Illinois 60637, USA}

\affiliation{$^2$Theoretical Division, Los Alamos National Laboratory, MS B213, Los Alamos, NM 87545, USA}

\date{\today}

\begin{abstract}
We compute the shear viscosity, $\eta$, at general temperatures $T$, in a BCS-BEC
crossover scheme which is demonstrably consistent with conservation
laws. The study of $\eta$ is important because it constrains microscopic
theories by revealing the
excitation spectra.
The onset of a normal state pairing gap
and the contribution from pair degrees
of freedom imply that $\eta$
at low $T$ becomes
small, rather than exhibiting the upturn predicted by
most others.
Using the local density approximation, we find quite reasonable agreement 
with just-published experiments.
\end{abstract}


\maketitle
The ultracold Fermi gases near the unitary limit are thought to
be related to quark-gluon plasmas \cite{ThomasViscosityScience_online}. Much attention
has focused on particle-physics-based calculations of the anomalously
low shear viscosity, $\eta$ \cite{Son1,Rupakvisc}. However, considerable
insight on the thermodynamics \cite{ThermoScience} and
various spectroscopic studies \cite{CSTL05,RFReview} has 
also been obtained via a condensed matter perspective.
This paper belongs to the second school in which
BCS theory is extended to accomodate arbitrarily strong interactions.
We apply
BCS to Bose Einstein condensation (BEC) crossover
theory \cite{CSTL05} to compute
$\eta$ demonstrating
consistency with central
sum rules \cite{KadanoffMartin} and conservation laws.

The shear viscosity is a powerful probe for testing microscopic
theories, because it reflects the
normal fluid component. As a result it is extremely sensitive
to the nature of the excitation spectrum.
In a low $T$ normal Fermi liquid phase with scattering
lifetime $\gamma^{-1}$ and effective mass $m^{\ast}$, 
$\eta=\frac{1}{5}nv^2_F\gamma^{-1} m^{\ast}$.
More generally, one can think of $\eta$ as characterized 
by the effective number of
the normal excitations ($n \rightarrow n_{eff}(T)$)
as well as their lifetime which we emphasize here is
a many body effect.
Crucial is an understanding of
how $n_{eff}$ depends on $T$.
BCS-based approaches have addressed the
behavior of the viscosity in a notable class of fermionic
superfluids-- helium-3
\cite{Viscositypapers,Dorfle80}.
Here experiments \cite{Helium3} indicate that $\eta$ 
drops off rapidly to zero in the superfluid phase, reflecting
the suppression of fermionic excitations at low $T$.
In
the helium-4 counterpart, the single particle bosonic
excitations couple to the collective (Nambu-Goldstone) modes,
leading to an upturn \cite{Woods} in $\eta$ at low $T$, which has
also been predicted (but not seen)
for the atomic Fermi superfluids \cite{Rupakvisc}.
In BCS-based superfluids, we stress that
Nambu-Goldstone
boson effects
do not naturally enter into the
transverse transport properties, such as $\eta$.

The Fermi superfluids involve \cite{Regge77} 
inter-dependent fermionic as well as bosonic (or pair) excitations,
here deriving from stronger-than-BCS attraction.  
In past literature there has been a focus on either one
\cite{Bruunviscosity} or the other \cite{Rupakvisc},
but not both.
Here we use a Kubo-based formalism which readily accomodates the
simultaneous bosonic and fermionic contributions and thereby
addresses $n_{eff}$
quite accurately
while the
alternative Boltzmann or kinetic theory-based approaches
do not naturally incorporate these multiple
statistical effects. 
The Kubo approach
includes scattering processes via
the lifetimes \cite{KadanoffMartin2} which appear in
the various Green's functions, while Boltzmann schemes treat
lifetimes via collision integrals.  However, because the physics of this
dissipation is principally associated with the many body
processes of boson-fermion inter-conversion,
it can be satisfactorily addressed
only in theories which treat the mixed statistics.
In this way it appears that Boltzmann based schemes may turn out to be
inadequate, particularly at low $T$.

Our central conclusion here is that both the effects of a
fermionic gap (with onset temperature $T^{\ast} > T_c$) and the
non-condensed pairs
act in concert to reduce $n_{eff}$ and thus
\textit{lower} the shear viscosity
at all $T < T^{\ast}$.
When quantitatively compared with
very recent shear viscosity experiments \cite{ThomasViscosityScience_online}
(we independently infer \cite{OurComparison} 
an estimated lifetime from
radio frequency data) the agreement is reasonable. 
Using previous thermodynamical
calculations \cite{ThermoScience}
of the trap energy $E$ and entropy density 
$s$, a plot of the trap-integrated $\eta$ decreases
with decreasing $E$, while $\eta/s$ appears to be roughly constant
at low $E$, but above the
universal quantum limit \cite{Son1}.  

We
compute viscosities using the
more
well controlled current-current correlation
functions \cite{KadanoffMartin},
$\tensor{\chi}_{JJ}$
rather than stress-tensor correlation
functions, via
%
$$\eta=-m^2\lim_{\omega\rightarrow0}\lim_{\mathbf{q}\rightarrow0}\frac{\omega}{q^2}\textrm{Im}\chi_T(\omega,\mathbf{q})$$
which is importantly constrained by the sum rule \cite{KadanoffMartin} 
\begin{equation}
\displaystyle{\lim_{\mathbf{q}\rightarrow0}}\displaystyle{\int_{-\infty}^{\infty}}\frac{d\omega}{\pi}\big(-\frac{\textrm{Im}\chi_T(\omega,\mathbf{q})}{\omega}\big)=\frac{n_n(T)}{m}, 
\label{rulechiT} 
\end{equation}
The transverse susceptibility
$\chi_T=(\sum_{\alpha=x}^z\chi^{\alpha\alpha}_{JJ}-\chi_L)/2$ with the longitudinal
$\chi_L=\hat{\mathbf{q}}\cdot\tensor{\chi}_{JJ}\cdot\hat{\mathbf{q}}$.
Here $n_n(T)$ corresponds to the number of particles
in the normal fluid and $n_n(T) \rightarrow n$
above
$T_c$, where
$n$ is the total particle
number.
Because dissipative transport in BCS-BEC theory is complex,
among the most persuasive checks  
on a proper characterization of the ``normal fluid''
is
consistency with sum rules \footnote{
Because it is more difficult to precisely enforce the counterpart longitudinal sum rule,
which also involve Goldstone boson effects, in
this paper we do not discuss the bulk viscosity.}.
Equation \eqref{rulechiT} is
more general and fundamental than sum rules derived in Ref.~\cite{TaylorRanderia}.

Our theoretical scheme is based on the BCS-Leggett ground state,
extended \cite{CSTL05} to non-zero temperature $T$. 
There are then two
contributions to the square of the pairing gap $\Delta^2 (T)
= \Delta_{\textrm{sc}}^2(T) + \Delta_{\textrm{pg}}^2(T)$, corresponding to
condensed (sc) and to non-condensed (pg) pairs, which are associated with
a pseudogap.
The fermions have dispersion
$E_{\mathbf{p}} \equiv \sqrt{ \xi_{\mathbf{p}}^2 + \Delta^2(T)  }$, where $\xi_{\mathbf{p}}=\epsilon_{\mathbf{p}}-\mu$ (In what follows, we omit the subscript $\mathbf{p}$ for 
convenience).

A consistent set of diagrams which satisfy Eq.(\ref{rulechiT}), include
\cite{CSTL05,Kosztin2,OurBraggPRL}, 
both Maki Thompson (MT) and two Aslamazov-Larkin (AL) diagrams.
In general, the
current-current correlation function is
$\tensor{\chi}_{JJ}=\tensor{P}+\frac{\tensor{n}}{m}+C_{J}$, where $C_{J}$ is 
known \cite{Kosztin2,OurBraggPRL} and
associated with excitations of the collective modes. 
All transport
expressions in this paper reduce to those of strict BCS theory
when the attraction is weak and $\Delta_{\textrm{pg}} =0$.

Consistency in linear response theory is based on the use of
Ward identities which connect transport to the fermionic self energy,
\begin{equation}
\Sigma({\bf p},\omega)
\equiv -i \gamma + \frac{\Delta_{pg}^2 }{\omega+\xi_{\textbf{p}}+i\gamma} +
\frac{\Delta_{sc}^2 }{\omega+\xi_{\textbf{p}}+i0^{+}}. 
\label{eq:2}
\end{equation}
Here we use a
broadened BCS self energy which was adopted experimentally in
Refs.~\cite{Jin6} and theoretically \cite{RFReview} to
address radio frequency (RF)
based studies in the cold gases. 
The condensed
pairs have the usual BCS self energy contribution, $\Sigma_{sc}$,
while the self energy of the non-condensed pairs $\Sigma_{pg}$
contains an additional damping term parameterized by $\gamma$.

To keep the equations simple and transparent we proceed in two stages.
We begin in the weak dissipation limit, by which we mean the lifetime
of the fermionic excitations is very long, so that to leading order
we may set $\gamma \approx 0^+$ in Eq.\eqref{eq:2}.
In this weak dissipation limit, using the MT and AL diagrams \cite{Kosztin2,OurBraggPRL,CSTL05} one can 
write more simply
\begin{eqnarray}
& &\tensor{P}(\omega,\mathbf{q})=\sum_{\mathbf{p}}\frac{\mathbf{p}\mathbf{p}}{m^2}\Big[\frac{E_++E_-}{E_+E_-}\big(1-f_+-f_-\big)\nonumber\\                                                                   & &\times\frac{E_+E_--\xi_+\xi_--\delta\Delta^2}{\omega^2-(E_++E_-)^2}-\frac{E_+-E_-}{E_+E_-}\nonumber\\      
& &\times\big(f_+-f_-\big)\frac{E_+E_-+\xi_+\xi_-+\delta\Delta^2}{\omega^2-(E_+-E_-)^2}\Big],
\label{eq:5}
\end{eqnarray}
where 
$\hbar=1$,
$E_{\pm}=E_{\mathbf{p}\pm\mathbf{q}/2}$, $f_{\pm}=f(E_{\pm})$ and $\delta\Delta^2=\Delta^2_{\textrm{sc}}-\Delta^2_{\textrm{pg}}$.

With these diagrams, one can prove 
consistency with the sum
rule for the transverse susceptibility (Eq.\eqref{rulechiT}). 
[For
the more general case the proof depends on the fact that
Eq.\eqref{rulechiT}
is closely
tied to the absence (above $T_c$) and presence (below $T_c$)
of a Meissner effect].
More explicitly, in the weak dissipation limit, we note
that the total number of particles can be written as
$n=\sum_{\mathbf{p}}\big(1-\frac{\xi}{E}
(1-2f)\big)$. 
The superfluid density at general temperatures is given by 
$n_s=m\mbox{Re}P^{xx}(0,0)+n = \frac{2}{3}\frac{\Delta^2_{\textrm{sc}}}{m}\sum_{\mathbf{p}}\frac{p^2}{E^2}\Big(\frac{1-2f}{2E}+\frac{\partial f}{\partial E}\Big)$.
The transverse susceptibility at $\mathbf{q}\rightarrow 0$ contains
no collective mode contributions; thus using
$\tensor{P}$
in  
Eq.\eqref{eq:5}, the left hand side of Eq.\eqref{rulechiT} is
$\sum_{\mathbf{p}}\frac{p^2}{6m^2}\Big[\frac{2\Delta^2_{\textrm{pg}}}{E^2}\frac{1-2f}{E}\nonumber\\
-4\frac{E^2-\Delta^2_{\textrm{pg}}}{E^2}\frac{\partial f}{\partial E}\Big]=\frac{n - n_s}{m} =  \frac{n_n}{m}$, thereby proving the sum rule.

We introduce the abbreviated notation
$\delta_1(\omega)=\delta(\omega-E_+-E_-)$, $\delta_2(\omega)=\delta(\omega-E_++E_-)$,
and $\theta$ is the polar angle.
Then based on
$\chi_T(\omega,\mathbf{q})$
the shear viscosity is   
\begin{eqnarray}
& &\eta=-m^2\lim_{\omega\rightarrow0}\lim_{\mathbf{q}\rightarrow0}\frac{\pi\omega}{2q^2}\sum_{\mathbf{p}}\frac{p^2\textrm{sin}^2\theta}{m^2}\Big[\big(1-f_+-f_-\big)\nonumber\\
&\times&\frac{E_+E_--\xi_+\xi_--\delta\Delta^2}{2E_+E_-}\big(\delta_1(\omega)-\delta_1(-\omega)\big)-(f_+-f_-)\nonumber\\
&\times&\frac{E_+E_-+\xi_+\xi_-+\delta\Delta^2}{2E_+E_-}\big(\delta_2(\omega)-\delta_2(-\omega)\big)\Big],
\label{eq:7}
\end{eqnarray}
If we now
take the low $\omega, q$ limits in
Eq.\eqref{eq:7},
$\eta$ assumes a form similar
to a stress tensor correlation
function.
We incorporate lifetime effects \cite{KadanoffMartin2}
(which preserve
the analytic sum rule consistency) by writing
$\delta(\omega\pm\mathbf{q}\cdot\nabla_{\mathbf{p}} E)=
\lim_{\gamma \rightarrow0}
\frac{\frac{1}{\pi}\gamma}{(\omega\pm\mathbf{q}\cdot\nabla_{\mathbf{p}} E)^2+
\gamma^2}$.
%
After this regularization
\begin{eqnarray}\label{eq:9}
\eta=\int_0^{\infty}dp\frac{p^6}{15\pi^2m^2}\frac{E^2-\Delta^2_
{\textrm{pg}}}{E^2}
\frac{\xi^2}{E^2}\Big(-\frac{\partial f}{\partial E}\Big)\frac{1}{\gamma}
\label{eq:6}
\end{eqnarray}
from which one can identify the effective carrier number
($n_{eff}(T) \propto  \eta \gamma $ discussed in the introduction, and verify that
it is dramatically suppressed at low $T$.

\begin{figure}
\includegraphics[width=3.4in,clip]
{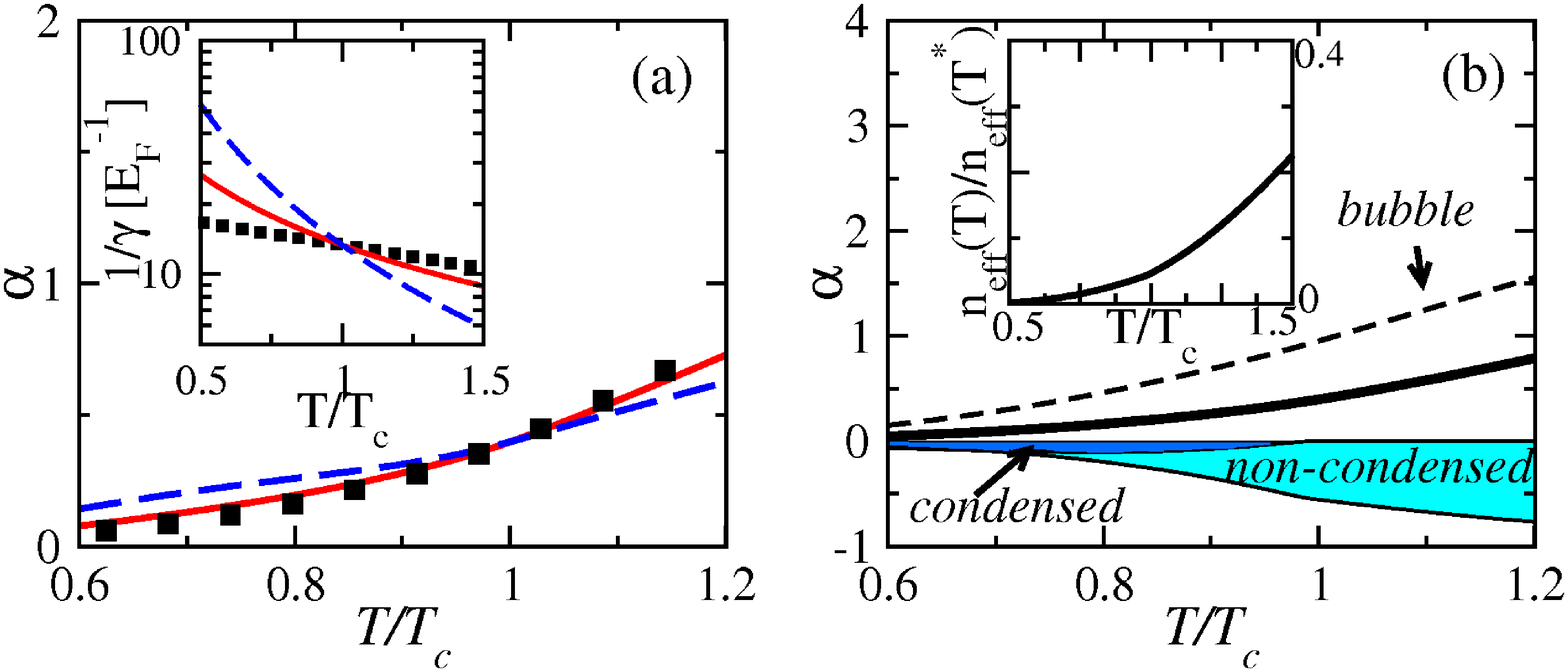}
\caption{(Color online) (a) Calculated (homogeneous) viscosity
$\eta=n\hbar\alpha$ for a unitary gas,
illustrating robustness
under changes in the value of the lifetime $1/\gamma$.
The color coded lifetimes
(in units $E_F^{-1}$)
indicated in the inset correspond to their counterparts in the main figure for
$1/\gamma$
deduced from fits to RF experiments (black squares) and both linear (red) and quadratic (blue)
dependences on $T/T_c$.
(b) Contributions to $\alpha$ (thick black curve) from condensed and non-condensed pairs and
from the simple bubble diagram. Inset plots the effective carrier number
as $ \propto \eta \gamma$, showing the decrease relative to high $T$.}
\label{fig:2}
\end{figure}

Physically, the two types of terms which appear in Eq.\eqref{eq:7}
are well known in standard BCS theory.
The first ($\delta_1$) refers to processes which require a minimal frequency of the
order of $2 \Delta(T)$; they arise from the contribution of fermions
which are effectively liberated by 
the breaking of pairs. The second of these
terms, involving $\delta_2$, arises from scattering of fermionic
quasi-particles and is the only surviving contribution to the viscosities,
which are defined in the 
$\omega\rightarrow0$ limit. 
Note that both contributions involve the \textit{difference} of the
condensed and non-condensed components   
($\delta\Delta^2=\Delta^2_{\textrm{sc}}-\Delta^2_{\textrm{pg}}$) with
opposite overall signs.
Importantly, the low $\omega$ quasi-particle scattering processes are
reduced by the presence of non-condensed pairs -- simply because more
pairs lead to fewer fermions.
By contrast in the
high $\omega \approx 2 \Delta $ limit
the number of contributing fermions will be increased
by the presence of non-condensed pairs when they are broken,
although this effect is only present in a $\omega \neq 0$ response.

Equation (\ref{eq:6}) is a generally familiar BCS expression 
\cite{Viscositypapers,Dorfle80}
except for the effects associated with
non-zero $\Delta_{\textrm{pg}}$ which appears as
a prefactor $1 - \frac{\Delta_{\textrm{pg}}^2}{E^2}$. 
This deviation from unity can be traced to the AL diagrams.
Note that all terms are reduced by a
coherence-effect-prefactor associated
with
$\xi^2/E^2$.
The fact that the non-condensed pairs suppress $\eta$
derives from the same physics discussed surrounding Eq.(\ref{eq:7})
and associated with the scattering term $\delta_1$.
The negative sign for this bosonic term comes physically from the fact that
when pairs are present there are fewer fermions to
contribute to the viscosity.
In a more formal sense, this contribution is required for
number conservation and the sum rule involving 
$\chi_T$.

Importantly, the expression in Eq.\eqref{eq:6}
can be generalized to the stronger dissipation limit, by introducing
generalized Green's functions defined in
Ref.~(\cite{ourqpi}), which represent
the various
$pg$ and $sc$ contributions. 
$\eta  =  -\sum_P\frac{2p^2_xp^2_y}{m^2}\Big[-(1+\frac{2\xi^2}{E^2})F_{
pg}^+F_{pg}^-                  
+G^+G^--F_{sc}^+F_{sc}^-
\Big]_{i\Omega_m\rightarrow0^+}$.
where $\pm$ is $P\pm Q/2$, $P=(\textbf{p},i\omega_n)$, $Q=(\textbf{q}\rightarrow 0,i\Omega_m)$, $\omega_n,\Omega_n$ are (fermionic, bosonic) Matsubara frequencies.
%
We emphasize that the
lifetime, $\gamma^{-1} $ is a temperature dependent many body effect
here found to be associated with
pair-fermion
inter-conversion, and partially quantified by the
cold atom Radio Frequency (RF) ``photoemission" 
experiments \cite{Jin6,RFReview} for $^{40}$K.
In $^6$Li, where the viscosity experiments are performed
\cite{ThomasVisc,ThomasJLTP08,ThomasViscosityScience_online}, 
we have previously estimated 
\cite{OurComparison}
this parameter by fitting
non-momentum-resolved RF experiments. We note
that
the introduction of this fermionic lifetime into a
broadened BCS form for the non-condensed pairs (or pseudogap contribution)
of Eq.\eqref{eq:2} is widely used in the high $T_c$ literature \cite{RFReview} as well.

In Figure.~\ref{fig:2}(a) we plot the shear viscosity for a unitary gas as computed
via the strong dissipation approach. We have verified that
the results are similar if we use the simpler form
of Eq.\eqref{eq:6} directly.
The plot is for $\alpha$ defined
as
$\eta \equiv \alpha
n\hbar$
versus
temperature for
a homogeneous system at unitarity and for a range of different lifetime
parameterizations. 
The inset to
Figure.~\ref{fig:2}(a) presents
a plot of the RF-deduced lifetime as black squares
along with a few alternative functional forms. Each of these corresponds
(via color coding) to the plots for $\alpha$ in the main body of the figure. 
In all cases $\alpha$ drops to zero at low temperatures, although one
can see that this is slightly countered by the fact that
the fermions are longer lived at low $T$. 
This figure should make it clear that the behavior shown here is quite
independent of any detailed models for the inverse lifetime $\gamma^{-1}$.
This largely follows from the intuition already embedded in
Eq.\eqref{eq:6} showing the dominant effect is due to
exponential decrease in the number of condensate excitations associated with
$s$-wave superfluidity.

In the inset of Figure.~\ref{fig:2}(b) 
we plot the effective carrier number defined as $\propto \eta \gamma$
as a function of $T$. The curve, normalized to the high temperature
value where $\Delta = 0$, shows a clear suppression of the carrier number
associated with the non-condensed pairs.
The main
body of the figure presents a breakdown of the various 
contributions to the viscosity
coming from 
this bubble term and from 
the contributions of condensed
(proportional to $\Delta_{sc}^2$) and non-condensed (proportional
to
$\Delta_{pg}^2$) pairs.

\begin{figure}
\includegraphics[width=3.5in,clip]
{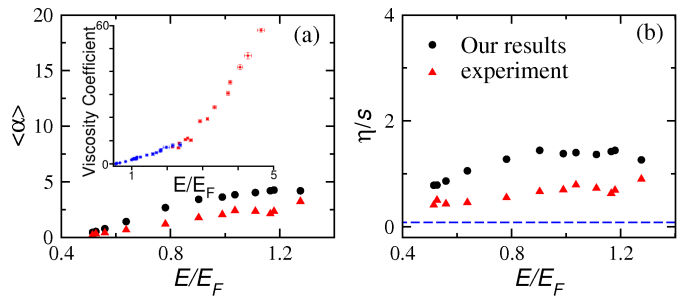}
\caption{ (Color online) (a) Comparison of shear viscosity 
$\eta \equiv \alpha
n\hbar$ 
and (preliminary) experiments \cite{ThomasJLTP08} (red triangles) at unitarity
for a trapped gas. In theory
plots (black dots) we use the
calculated thermodynamics for the trap energy $E$ and entropy density
$s$. 
The inset in (a) plots data from \cite{ThomasViscosityScience_online}. 
(b) Comparison of $\eta/s$. The blue dashed line labels the quantum lower limit of $\eta/s$ given by Ref.~\cite{Son1}.
}
\label{fig:3}
\end{figure}

We turn now to calculations in a trap based on the local density approximation.
The inset to
Figure.~\ref{fig:3}(a) presents a plot of experimental data from Ref.~\cite{ThomasViscosityScience_online} 
The main body of
Figure.~\ref{fig:3}(a)
presents a comparison of the viscosity coefficient
$\alpha$ between theory (based on the RF-deduced lifetime), as black dots, and experiment \cite{ThomasJLTP08} (red triangles)
as a function
of $E$. 
Figure.~\ref{fig:3} 
(b) shows the comparison of $\eta/s$ where $s$ is the entropy density.
We find
that $\eta/s$ appears to be relatively $T$ independent at the lower
temperatures.
Here we have used the calculated trap thermodynamics 
\cite{ThermoScience} to rescale the various axes,
and our calculations are based on 
the same trap averaging procedure as in Ref.~\cite{ThomasJLTP08}.
One can anticipate that, particularly at the lower $T$, the trap-integrated viscosity will be artificially higher
than for the homogeneous case, since $\eta$ will be dominated by 
unpaired fermions at the trap edge. 
Overall, it can be seen that our                                              calculations agree favorably with the preliminary experimental 
data shown in the inset and figure.
Interestingly, the observed behavior appears more consistent with
previous helium-3 experiments
\cite{Helium3}
than those in helium-4 \cite{Woods}.

There appear to be no other BCS-BEC based
calculations of $\eta$ in the literature
which address the entire range of $T$ below $T_c$ and also
the consistency check of 
Eq.\eqref{rulechiT}.
Bruun and Smith \cite{Bruunviscosity}
have studied the above $T_c$ shear viscosity 
and 
recognized \cite{Bruunviscosity}
that the pseudogap
reduces $\eta$. However, the diagram set which was used was
``not conserving''\cite{Bruunviscosity}. 
Rupak and Schafer \cite{Rupakvisc}
argued that $\eta$ is 
dominated
by the Goldstone bosons or phonons and predicted an upturn at
the lowest $T$ in
both $\eta$ and
$\eta/s$.
This upturn has not yet been seen experimentally, and it
seems also be missing in earlier Boltzmann-based
approaches \cite{Shahzamanian02}.
Importantly, in BCS-based fermionic superfluids the Goldstone bosons
do not couple to a transverse response such as $\eta$.

In summary,
in this paper we have effectively ascribed the low viscosity
of the strongly interacting Fermi gases to effects associated
with the depression in the normal fluid carrier density
($n_{eff}(T))$ arising from non-condensed pairs and the related pseudogap. Our
viscosity decreases monotonically as $ T \rightarrow 0$,  as seen in superfluid
helium-3 
\cite{Helium3}
and in very recent Fermi gas data \cite{ThomasViscosityScience_online}, suggesting the
absence of a ground state normal fluid. For
these fermionic superfluids, lifetime effects are associated
with fermion-boson interconversion. Moreover because of this
mixed statistics \cite{Regge77}, Boltzmann approaches are not
amenable to including these lifetime contributions.
We emphasize that all approaches to transport must necessarily
demonstrate, as we do here, consistency with number conservation.

With these on-going debates and the possible impact for other
physics sub-disciplines,  measurements and theories of
the shear viscosity take on particular importance. As shown here they
constrain microscopic theories through
the
detailed information they provide
about the nature of the superfluid excitation spectrum.
Interestingly, calculations of the $\omega \rightarrow 0$ conductivity
lead, by analogy,
to ``bad metal" behavior \cite{Emery} in the cuprates, which may be a counterpart
to perfect fluidity in the cold gases.

This work is supported by NSF-MRSEC Grant
0820054. We thank Ben Fregoso and also
Le Luo and John Thomas for helpful
conversations. 
C.C.C. acknowledges the support of the U.S.
Department of Energy through the LANL/LDRD Program.

\bibliographystyle{apsrev}  

\end{document}